\documentclass{article}

\usepackage{microtype}
\usepackage{graphicx}
\usepackage{subfigure}
\usepackage{booktabs} 

\usepackage{hyperref}

\usepackage[accepted]{icml2025}

\usepackage{amsmath}
\usepackage{amssymb}
\usepackage{mathtools}
\usepackage{amsthm}

\usepackage[capitalize,noabbrev]{cleveref}

\usepackage{multirow, makecell} 
\definecolor{mygreen}{RGB}{67, 179, 86} 
\definecolor{myyellow}{RGB}{247, 213, 4}
\definecolor{myblue}{RGB}{41, 182, 205}
\definecolor{tableyellow}{RGB}{253, 236, 131}
\definecolor{tablegreen}{RGB}{212, 239, 201}
\definecolor{tableblue}{RGB}{197, 236, 243}
\definecolor{Gblue}{RGB}{85, 168, 220}
\definecolor{Gred}{RGB}{232, 71, 96}
\definecolor{Gyellow}{RGB}{255, 192, 0}
\definecolor{Ggreen}{RGB}{108, 186, 75}

\theoremstyle{plain}

\theoremstyle{definition}

\theoremstyle{remark}

\usepackage[textsize=tiny]{todonotes}

\icmltitlerunning{FLOWER: Flow-Based Estimated Gaussian Guidance for General Speech Restoration}

\begin{document}

\twocolumn[
\icmltitle{FLOWER: Flow-Based Estimated Gaussian Guidance \\ for General Speech Restoration}



\icmlsetsymbol{equal}{*}

\begin{icmlauthorlist}
\icmlauthor{Da-Hee Yang}{equal,yyy}
\icmlauthor{Jaeuk Lee}{equal,yyy}
\icmlauthor{Joon-Hyuk Chang}{yyy}
\end{icmlauthorlist}

\icmlaffiliation{yyy}{Department of Electronic Engineering, Hanyang University, Seoul, South Korea}

\icmlcorrespondingauthor{Joon-Hyuk Chang}{jchang@hanyang.ac.kr}


\vskip 0.3in
]



\printAffiliationsAndNotice{\icmlEqualContribution} 

\begin{abstract}
We introduce FLOWER, a novel conditioning method designed for speech restoration that integrates Gaussian guidance into generative frameworks. By transforming clean speech into a predefined prior distribution (e.g., Gaussian distribution) using a normalizing flow network, FLOWER extracts critical information to guide generative models. This guidance is incorporated into each block of the generative network, enabling precise restoration control. Experimental results demonstrate the effectiveness of FLOWER in improving performance across various general speech restoration tasks.
\end{abstract}

\section{Introduction}
General speech restoration (GSR) aims to enhance speech quality and intelligibility in real-life environments, where signals are often degraded by multiple distortions such as noise, reverberation, and bandwidth degradation. Traditional methods, including denoising, dereverberation, and bandwidth extension (BWE), primarily address these distortions in isolation \cite{dnnse, dnnbwe, dnnreverb, nuwave, cdiffu, sgmse, diffBWE}. However, in real-world scenarios, multiple distortions frequently occur simultaneously, posing greater challenges for effective restoration.

Recent studies have demonstrated progress in addressing multiple distortions, highlighting the potential of generative models for GSR. \citep{voicefixer, universal, usee, universepp}.
Among generative models, diffusion-based models \cite{SDE, DDPM} have shown notable promise in producing high-quality, natural-sounding speech \cite{cdiffu, sgmse+}. Recently, these models integrate conditioning mechanisms to improve performance by utilizing complex network architecture or task-specific condition information \cite{universal, usee, universepp}.
For example, discriminative-based conditioning networks extract enhanced features from distorted speech and integrate them into generative models \cite{universal, universepp}.
Similarly, task-specific condition information, such as signal-to-noise ratio (SNR) or environmental details, is injected to provide context for specific restoration tasks \cite{usee}.

Although these approaches enhance diffusion models, they often rely on deterministic outputs from discriminative networks or task-specific conditions, which may not fully encapsulate the diverse complexities of speech restoration tasks. 
This dependency on predefined conditions and features limits their flexibility and scalability.

To address these limitations, we propose FLOWER (FLOW-based Estimated Gaussian guidance for general speech Restoration), a novel conditioning approach that introduces Gaussian guidance as a conditioning feature within generative models.
This guidance is derived from a normalizing flow (NF) network \cite{Glow}, which is specifically designed to extract residual information that bridges the gap between clean speech and the latent features of the generative model.
Unlike deterministic outputs, the Gaussian guidance encapsulates oracle knowledge of clean speech in a stochastic form, enabling generative models to better navigate the restoration process.

FLOWER integrates Gaussian guidance into each block of the diffusion model, enabling dynamic adjustments to address varying levels of distortions during the generation of enhanced speech. By leveraging this fine-grained conditioning, the model generates data distributions that are more closely aligned with clean speech, addressing multiple distortions more effectively. Furthermore, to further improve sampling efficiency, we extend FLOWER to a flow-matching (FM) \cite{FM} model using an optimal transport (OT) path. We also introduce a time-adaptive scaling mechanism for Gaussian guidance, which dynamically adjusts the conditioning influence as the diffusion process progresses, ensuring effective utilization of condition information.

The main contributions of FLOWER are as follows:

\noindent 1. \textbf{Novel Conditioning Approach}: We present FLOWER, which extracts Gaussian guidance from an NF network to generate residual information distributed in a prior distribution (e.g., Gaussian). This Gaussian guidance provides critical knowledge that is absent from the latent features of the generative model, enhancing its restoration capabilities.

\noindent 2. \textbf{Efficient Inference}: FLOWER extracts Gaussian guidance from the NF network during training. However, during inference, this guidance is sampled directly from a Gaussian distribution, eliminating the need for the NF network and ensuring computational efficiency. 

\noindent 3. \textbf{Performance Improvements}: FLOWER demonstrates superior performance in general speech restoration tasks, surpassing baseline diffusion models in metrics related to noise reduction, dereverberation, and bandwidth extension. FLOWER also achieves higher sampling efficiency through the FM model with time-adaptive scaling.

\section{Related Work}
\subsection{General Speech Restoration}
Speech enhancement aims to restore intelligibility and quality in distorted speech signals. Traditional approaches have primarily addressed individual distortions, focusing on tasks like denoising, dereverberation, or BWE \cite{dnnse, dnnbwe, dnnreverb, nuwave, cdiffu, sgmse, diffBWE}. However, real-world environments present challenges where distortions occur simultaneously, making single-distortion models insufficient for comprehensive restoration.
Recognizing this limitation, recent research has shifted towards general speech restoration (GSR), a task aimed at simultaneously addressing multiple distortions such as noise, reverberation, and bandwidth degradation. 

VoiceFixer\cite{voicefixer} was among the first to define this task explicitly, introducing a framework that combines task-specific models within an analysis-and-synthesis structure, using a residual U-Net to restore mel-spectrograms followed by a vocoder to generate waveforms. Similarly, HD-DEMUCS \cite{hddemucs} employed a parallel decoder architecture to simultaneously enhance and reconstruct speech, leveraging the strengths of the DEMUCS model \cite{demucs} to effectively handle multiple distortions.

While deterministic deep learning models have achieved remarkable results, their mappings from noisy to clean speech often lack the flexibility and quality required for more complex scenarios. Generative models, particularly diffusion models, have demonstrated the ability to generate high-quality and natural-sounding speech, making them well-suited for GSR tasks.
Diffusion-based speech enhancement models such as CDiffuSE \cite{cdiffu} and SGMSE+ \cite{sgmse+} highlight the potential of leveraging diffusion models for tasks like denoising, dereverberation. Building on this foundation, UNIVERSE \cite{universal} introduced a conditioning network designed to extract enhanced features from distorted speech, integrating these features into a score-based diffusion model via joint training. The recently proposed UNIVERSE++ \cite{universepp} further advanced this concept by incorporating adversarial loss for enhanced restoration performance.

In line with these advancements, we propose a novel conditioning approach for diffusion-based GSR called FLOWER. Unlike prior methods that rely on enhanced features or task-specific conditions, FLOWER introduces Gaussian guidance, enabling it to handle multiple distortions more effectively. By integrating FLOWER into the generative process, we aim to demonstrate significant improvements in both performance and efficiency for GSR. The subsequent sections detail how we extract and incorporate Gaussian guidance into the generative framework, highlighting the unique contributions of our approach.

\subsection{Problem Formulation}
We aim to address GSR in adverse environments characterized by noise, reverberation, and bandwidth degradation. In our scenario, the observed speech signal is a distorted signal of clean speech $x$ with a room impulse response $r$ and background noise $n$, represented as $y=h(x*r)+n$. Here, $*$ denotes the convolution operation and the function $h$ introduces spectral distortions through low-pass filtering. The objective of this task is to estimate an enhanced speech signal $\hat{x}$ that closely approximates the original clean speech $x$ while mitigating the effects of distortions.

\section{Preliminaries}
In this work, we build upon two key generative frameworks: score-based diffusion models and flow-matching models. These frameworks provide the foundation for generating data distributions from noise or prior distributions. To enhance their capabilities, we propose the integration of Gaussian guidance, a novel conditioning feature extracted from a conditional normalizing flow network. 

This section introduces the generative frameworks underlying our approach. First, we detail score-based diffusion models that leverage score matching to iteratively refine noise into data. Next, we explain flow-matching models that estimate probability paths between data and prior distributions through vector fields. Finally, we describe the conditional NF network, which extracts Gaussian guidance by mapping input data to a normalized prior distribution.

\subsection{Generative Frameworks}
\subsubsection{Score-based diffusion model.}
A diffusion model is an advanced generative model that iteratively refines a noise distribution to generate a data distribution. The diffusion model operates in two processes: forward and reverse. The forward process transforms the data distribution into a prior distribution (e.g., Gaussian distribution), while the reverse process gradually removes noise using a sampling method (e.g., Langevin dynamics) that generates data from the prior distribution. Until recently, diffusion models have been divided into score-based \cite{score} and denoising diffusion probabilistic models \cite{DDPM}. \cite{SDE} introduces a unified approach based on a stochastic differential equation (SDE) that satisfies the probability trajectories of score-based and denoising diffusion probability models. The SDE-based forward process is defined as follows:
\begin{equation}
     dx_t=f(x_t,t)dt+g(t)dw_t,
\end{equation}
where $w_t$ represents the standard Wiener process and $f(\cdot, t)$ and $g(t)$ denote the drift and diffusion coefficients, respectively, which follow a predefined scheduler according to $t\in[0, 1]$. 
Through the forward process, the data distribution $x_0$ is gradually converged into the prior distribution $x_1$. 
According to \cite{RP}, the reverse process corresponding to the forward process can be obtained based on the SDE as follows:
\begin{equation}
     dx_t=[f(x_t,t)-g(t)^2\nabla_{x_t}\log p(x_t)]dt+g(t)d\bar{w}_t,
\end{equation}
where $\bar{w}_t$ denotes the backward standard Wiener process, and $\nabla_{x_t}\log p(x_t)$ represents the score (gradients of the log probability density) of the distribution $p(x_t)$.
In this reverse process, the noise in $x_1$ is gradually removed, ultimately generating $x_0$. 
By applying the score $\nabla_{x_t}\log p(x_t)$ to the reverse process, data can be generated from the prior distribution, as described in Eq. (2).
To enable this, we train a score-based diffusion model to predict the score corresponding to input $x_t$ using the following objective:
\begin{equation}
     ||s_{\theta}(x_t,t)-\nabla_{x_t}\log p(x_t|x_0)||_2^2,
\end{equation}
where $p(x_t|x_0)$ denotes the transition kernel, indicating the distribution of $x_t$ conditioned on $x_0$. The function $s_{\theta}(\cdot,t)$ is a neural network designed to predict the score, and this training is known as score matching.

\subsubsection{FM-based model using OT path.}
FM \cite{FM} estimates the probability path between $x_1$ and $x_0$, which are sampled from the data and prior distributions, respectively. Assuming that a function $f$ transforms $x_0$ into $x_t$ at time $t\in[0,1]$, this transformation can be expressed as $f(x_0,t)=x_t$. Consequently, the ordinary differential equation (ODE) for $x_t$ is defined as follows:
\begin{equation}
     \frac{df}{dt}=\frac{dx_t}{dt}=u_t,
\end{equation}
where $u_t$ denotes the vector field responsible for generating $x_t$ from $x_0$. To estimate $u_t$, we train a neural network $v_t(\cdot,t)$ using the following objective:
\begin{equation}
     ||v_t(x_t,t)-u_t||_2^2.
\end{equation}
This training process is referred to as FM, where $u_t$ should be defined. If $x_t$ ($t\in[0,1]$) is distributed in Gaussian, then $f$ can be represented by an affine transformation:
\begin{equation}
     f(x_0,t)=\sigma_tx_0+\mu_t.
\end{equation}
The specific probability path depends on $\mu_t$ and $\sigma_t$, which significantly influence the sampling efficiency. To define this probability path, $\mu_t$ and $\sigma_t$ must satisfy two conditions: (i) $\mu_1$ and $\sigma_1$ correspond to $x_1$ and $\sigma_{min}(\cong0)$, respectively, and (ii) $\mu_0$ and $\sigma_0$ are set to 0 and 1, respectively. Among the paths that satisfy these conditions, \cite{FM} proposed an Optimal Transport (OT) path that changes $x_t$ linearly over time. 

The OT path is the simplest probability path that satisfies the above conditions, with a time-dependent probability density defined as follows:$\mu_t=tx_1$ and $\sigma_t=1-(1-\sigma_{min})t$.
Therefore, the affine transformation along the OT path is given by $f(x_0,t)=tx_1+(1-(1-\sigma_{min})t)x_0$.
As a result, the vector field $u_t$ is defined as follows: 
\begin{equation}
     u_t=x_1-(1-\sigma_{min})x_0.
\end{equation}
Since the formula for $u_t$ does not depend on time $t$, the OT path linearly transforms $x_0$ into $x_1$. Consequently, FM using the OT path enhances sampling efficiency compared to the score-based diffusion model.

For inference, we use an ODE solver based on Euler steps to sample from the FM model. Specifically, the sampling steps are defined as follows:
\begin{equation}
x_{t+\frac{1}{N}} := x_t + \frac{1}{N} v_\theta(x_t, y, z, t),
\quad t := t + \frac{1}{N},
\end{equation}
where $N$ represents the total number of time steps, and $v_{\theta}(\cdot)$ is the vector field predicted by the FM model. The use of Euler steps provides a straightforward and efficient method for solving the ODE, ensuring stable and reproducible sampling results.

\begin{figure}[t!]
\centering
\includegraphics[height=2.5cm, width=\linewidth]{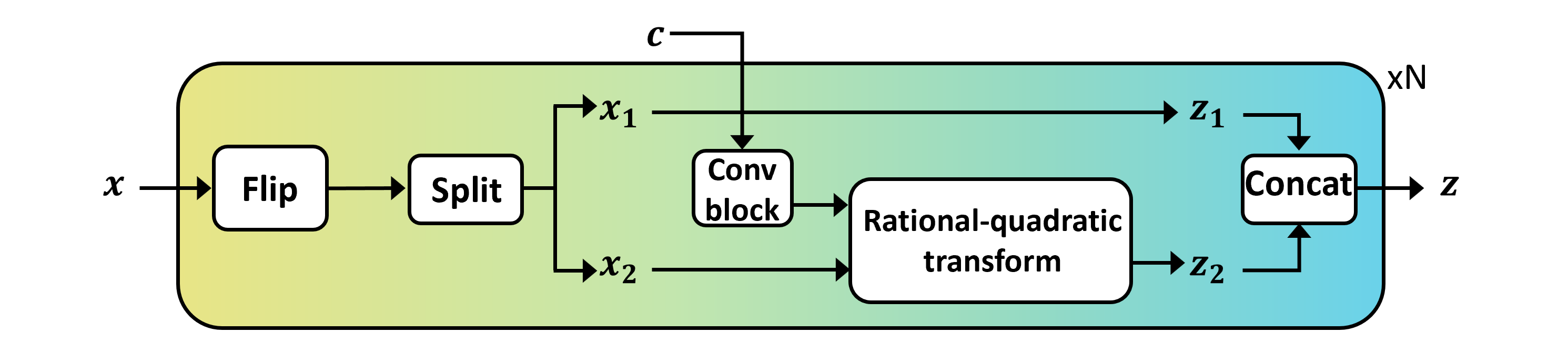}
\caption{Normalizing flow model architecture based on rational-quadratic transform. $x$, $c$, and $z$ are data (clean speech), a conditional feature (latent representation of diffusion model), and Gaussian noise, respectively. Conv block consists of dilated convolutions, and the number of block $N$ is 4.}
\end{figure}

\subsection{Conditional Normalizing Flow Network}
A Normalizing flow (NF) network is a type of generative model that uses an inverse function of flow to generate data.
Inspired by \cite{N2N}, we construct an NF network to extract the latent variable $z$ from the conditional NF network, as illustrated in Figure 1.
This conditional NF network takes two inputs: $x$ and $c$, to generate a conditional data distribution $p(x|c)$ that is normalized to the prior distribution.
The log-likelihood of the data distribution $p(x|c)$ is calculated as follows:
\begin{equation}
     \log p(x|c)=\log p(z|c)+\log|\det(\frac{\partial f(x)}{\partial x})|,
\end{equation}
where $p(z|c)$ represents the output of the NF network.
To train the NF network, the negative log-likelihood $-\log p(x|c)$ is decomposed into Kullback-Leibler (KL) divergence and entropy as follows:
\begin{equation}
     \text{KL}(p(z|c)|q(z))+H(x|c),
\end{equation}
where $q(z)$ is the prior distribution (typically standard Gaussian), and $H(x|c)$ denotes the constant data entropy. According to \cite{disentangle}, minimizing $\text{KL}(p(z|c)|q(z))$ reduces the mutual information between $z$ and $c$, effectively disentangling $z$ from $c$. This allows us to extract residual information $z$ from $x$, independently of $c$, since $q(z)$ is selected independently of $c$.
The mutual information $I(z,c)$ between $z$ and $c$ is represented by:
\begin{multline}
     I(z,c) = \int p(z,c) \log \frac{p(z,c)}{p(z)p(c)} = \int p(z,c) \log \frac{p(z|c)}{p(z)} \\ 
     = \int p(z,c) \log p(z|c) - \int p(z) \log p(z) \\
     \leq \int p(z,c) \log \frac{p(z|c)}{q(z)} = KL(p(z|c) \| q(z)).
\end{multline}
In our approach, we use clean speech as $x$ and the latent feature of the diffusion model as $c$ to extract residual information $z$ from the clean speech. This $z$ is then used as a conditioning feature for the diffusion model.

\begin{figure*}
  \centering
  \includegraphics[height=5.6cm, width=0.96\linewidth]{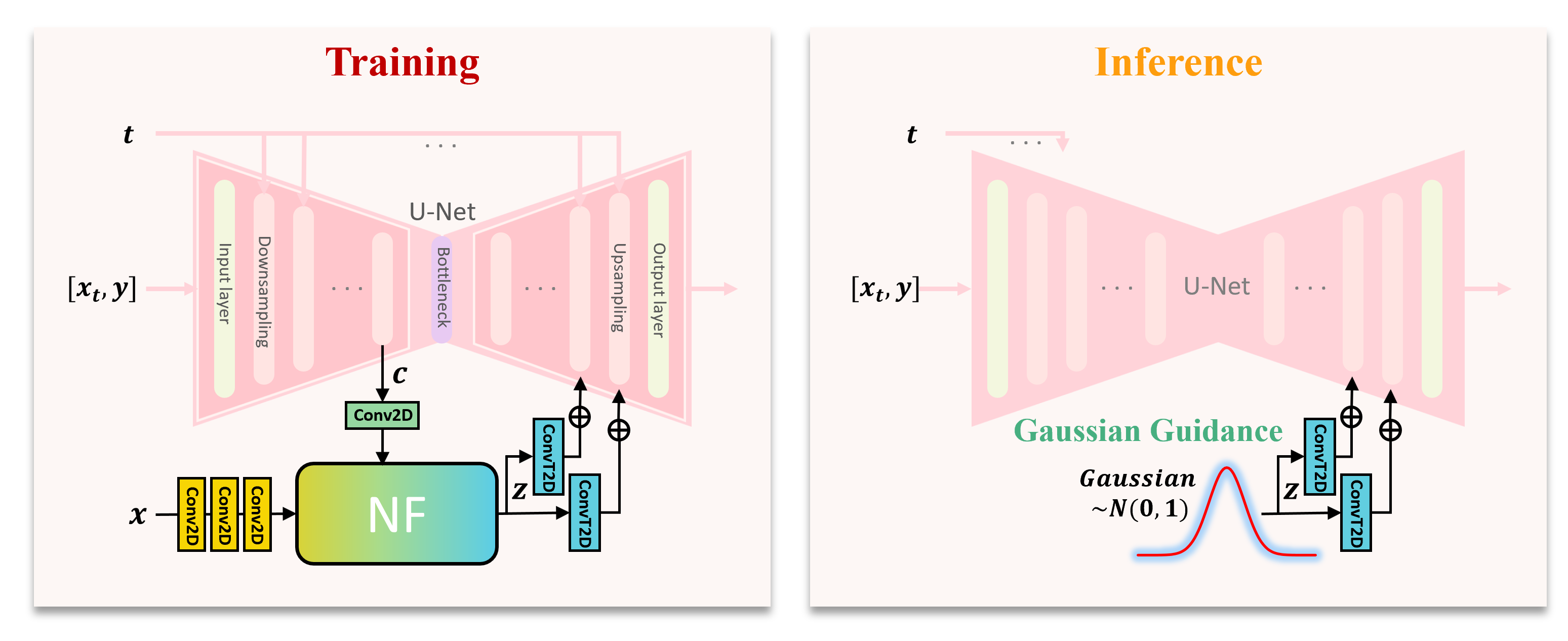}
  \caption{The overall architecture of FLOWER approach. Gaussian guidance is extracted from the NF network during the training (a), but it is extracted from a Gaussian distribution during the inference (b). In the training process, the latent feature $c$ passes through a \textcolor{mygreen}{Conv2D layer (256, 1, 1, 1)}, and the input speech spectrogram passes through three \textcolor{myyellow}{Conv2D layers}, each \textcolor{myyellow}{(1, 1, 4, 4)}. Subsequently, the output $z$ of the NF network passes through a \textcolor{myblue}{ConvTranspose 2D layer (1, 256, 32, 32)} and \textcolor{myblue}{(1, 128, 64, 64)} respectively, before being added to the diffusion network. Each shape represents (input channels, output channels, kernel size, and stride).}
\end{figure*}

\section{Method}
\subsection{Objective}
The primary objective of the FLOWER approach is to enhance the ability of speech restoration by introducing a novel conditioning feature that implicitly encompasses oracle knowledge of clean speech. This is achieved by incorporating the output of an NF network, termed Gaussian guidance, into generative frameworks. By leveraging this advanced conditioning, the FLOWER approach aims to more accurately predict a data distribution that closely aligns with the clean speech distribution, thereby improving the effectiveness of speech restoration.

\subsection{FLOWER Architecture}
The overall architecture of the proposed FLOWER approach, depicted in Figure 2, integrates an NF network with a diffusion model to provide an estimated Gaussian guidance as a conditioning feature. The NF network generates a Gaussian-distributed latent variable $z$ that is then utilized by the diffusion model as a structured and contextually relevant conditioning feature to enhance its generative and restoration capabilities.

During training, the NF network processes clean speech and latent features from the diffusion model to produce the Gaussian guidance $z$. This output is deterministic, as it is specifically shaped by the input $x$ (clean speech) and the condition $c$ (latent features). Although $z$ resides within a Gaussian distribution, it is not Gaussian noise; rather, it encapsulates statistical characteristics of clean speech, guided by the inputs. This guidance is then injected into the U-Net blocks, enabling fine-grained control over the restoration process and improving generalization.

During inference, the NF network is bypassed, and the Gaussian guidance $z$ is directly sampled from a Gaussian distribution. Unlike white noise, which is purely random and lacks context, this sampled guidance reflects the statistical characteristics learned during training. This structured information enables the diffusion model to maintain effective conditioning and produce high-quality restored speech even without the NF network.
This transition occurs because the NF network is trained to approximate a Gaussian distribution, allowing the model to maintain effective conditioning during inference \cite{VF}.
The following subsections detail the components and the process of extracting Gaussian guidance.

\subsubsection{Generative Network.} 
The backbone of the generative model is a multi-resolution U-Net structure \cite{unet}, as utilized in \cite{sgmse+}.
This network is designed to process complex spectrograms, estimating both real and imaginary components of the input signals.
It takes three inputs: a diffused signal ($x_t$), a distorted input ($y$), and time-step information ($t$), as shown in Figure 2.
The architecture consists of several stages: an input layer, downsampling layers, a bottleneck, upsampling layers, and an output layer.
The diffused and distorted signals are concatenated and processed through these layers, with the time-step information embedded into each block to capture time-dependent features.

The generative network ultimately estimates either a score for score-matching or a vector field for flow-matching, guiding the data distribution towards the target clean speech distribution.
Since the outputs depend on both the input and conditioning features, the generative model can generate a distribution that is closer to the target data distribution by providing relevant information as a conditioning feature.
To achieve this, we leverage the NF network to extract useful information from clean speech, which is then used to refine the output of the generative network.

\subsubsection{NF Network.}
To generate condition information, we use an NF network based on rational-quadratic transform \cite{NSF} as illustrated in Figure 1. The NF network takes two inputs, $x$ and $c$. In Figure 1, $x_1$, parts of $x$, and $c$ are the inputs of the Conv block and are added element-wise. The Conv block consists of 3 dilated convolutions with a dilation of $[0,3,9]$. In addition, the Conv block estimates some coefficients for the rational-quadratic transform that makes $z_2$ from $x_2$. Here, $z$ and $c$ are disentangled as referenced in \cite{disentangle}.

In our approach, the NF network takes clean speech ($x$) and the latent representation ($c$) from the last downsampling layer of the generative network, as shown in Figure 2.
The NF network then generates residual information $z$ that contains knowledge derived from the clean speech $x$ - knowledge that cannot be obtained from the latent representation of the diffusion network alone.
As depicted in Figure 2, the residual information $z$ (Gaussian guidance) is projected by the ConvTranspose 2D layers and injected into the last two upsampling layers of the generative network via element-wise summation.
The loss function for the FLOWER architecture combines the generative network loss ($\mathcal{L}_{UNet}$) and the NF network loss ($\mathcal{L}_{NF}$): $\mathcal{L}_{FLOWER} = \mathcal{L}_{UNet} + \mathcal{L}_{NF}$. 
The U-Net loss measures the mean squared error loss between the output of the backbone network and the target value, as described by Eqs. (3) and (5) for the score- and flow-matching models, respectively, while the NF model loss is defined as in Eq. (9).

\subsection{The Effect of FLOWER Approach}
\subsubsection{Estimated Gaussian Guidance.}
Effective guidance strategies for diffusion models have gained increasing attention because of their ability to effectively adapt to various speech restoration tasks \cite{universal, usee, universepp}.
However, conventional approaches rely on task-specific condition information that can limit the solution space.
Our approach introduces a novel form of conditioning information: Gaussian guidance estimated from the NF network.
During training as shown in Figure 2, this Gaussian guidance is aligned with a predefined prior distribution (e.g., Gaussian distribution) and provides the generative model with conditioning information that conveys oracle knowledge.
This allows for precise adjustments and enhancements in generating the data distribution, ultimately guiding the diffusion model towards optimal generative performance.

A key advantage of the FLOWER approach is that during inference, Gaussian guidance is directly sampled from a Gaussian distribution rather than being generated by the NF network. This eliminates the need for a conditioning network during inference, distinguishing our approach from conventional methods that require additional networks to provide conditioning information. By encapsulating clean speech information within the Gaussian guidance, our method effectively transmits oracle knowledge and enhances the generative capabilities of the model.

\subsubsection{Time-adaptive FLOWER on FM.}
FM with the OT path offers a more efficient and shorter probability path from the prior distribution to the data distribution. To enhance sampling efficiency, we apply the FM framework based on the OT path to the backbone network instead of score-based diffusion (SGMSE+), naming it FGMSE+. While SGMSE+ and FGMSE+ share the same architecture, the loss function and sampling strategy differ, with FGMSE+ utilizing flow-matching instead of score-matching.

To further improve the effectiveness of FGMSE+, we introduce a time-adaptive Gaussian guidance mechanism. As the time step $t$ increases, the input $x_t$ gradually approaches the clean speech distribution. Our method scales the Gaussian guidance according to the time step, making the guidance less influential as $x_t$ becomes closer to clean speech. Specifically, we scale the condition information by a factor of $1 - t$, corresponding to the narrowing gap between the clean speech and $x_t$ distributions.
This time-adaptive approach allows the Gaussian guidance to dynamically adjust to the level of distortion in the latent features, ensuring that the diffusion model effectively restores clean speech across different time steps. By adjusting the influence of the guidance according to the progression of the model, we significantly enhance the performance and efficiency of the generative process.
\begin{figure}
\centering
\includegraphics[height=6.5cm, width=\linewidth]{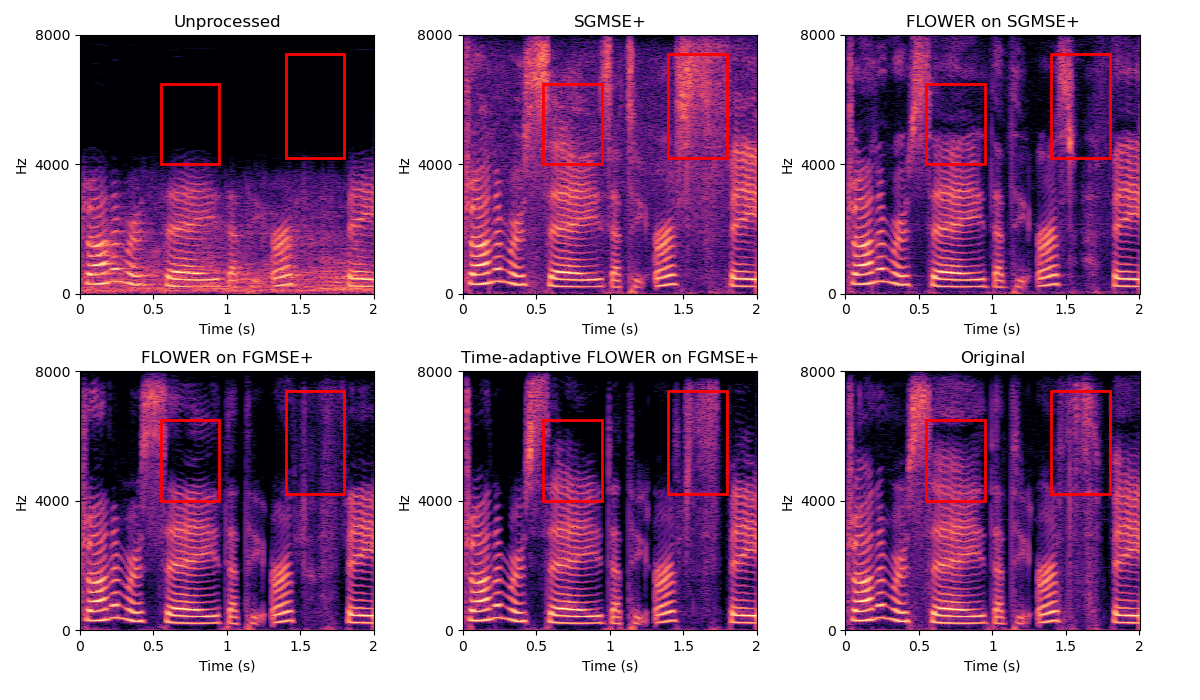}
\caption{Comparison between the restored spectrograms on ``matched" scenario.}
\end{figure}

\begin{table*}[]
\caption{Performance comparison between the baseline SGMSE+ and the proposed FLOWER models at 20 sampling steps. ``Unprocessed” represents the distorted speech signal by multiple distortions. ``SGMSE+” represents the main baseline model, while “FLOWER” denotes the proposed model applying our conditioning method to SGMSE+. ``Matched” scenarios refer to the multi-distortion test sets from WSJ+CHiME4, while ``Mismatched” scenarios refer to the multi-distortion test sets from VCTK+DEMAND. Higher scores indicate better performance.}
\centering
\begin{tabular}{cccccccc}
\hline \toprule
Method      & Scenarios                & PESQ ($\uparrow$) & CSIG ($\uparrow$) & CBAK  ($\uparrow$) & COVL ($\uparrow$) & SRMR ($\uparrow$) & SISDR ($\uparrow$)   \\ \midrule
Unprocessed & \multirow{3}{*}{Matched}   & 1.61 & 1.80 & 2.21 & 1.65 & 3.96 & 2.54 \\ \addlinespace[0.7mm]
SGMSE+       &                              & 2.10 & 3.52 & 2.78 & 2.87 & 6.08 & 7.14    \\ \addlinespace[0.7mm]
 \textbf{FLOWER}      &                              & \textbf{2.23} & \textbf{3.62} & \textbf{2.90} & \textbf{3.00} & \textbf{6.61} & \textbf{8.05}    \\ \addlinespace[0.4mm] \hline \hline \addlinespace[0.1cm]
Unprocessed & \multirow{3}{*}{Mismatched} & 1.78 & 1.80 & 2.25 & 1.73 & 5.91 & 2.58    \\ \addlinespace[0.7mm]
SGMSE+       &                              & 1.82 & 2.83 & 2.70 & 2.37 & 8.57 & 6.63    \\ \addlinespace[0.7mm]
 \textbf{FLOWER}      &                              & \textbf{1.95} & \textbf{2.99} & \textbf{2.81} & \textbf{2.51} & \textbf{9.34} & \textbf{7.55}    \\ \addlinespace[0.4mm] \Xhline{1.1pt}
\end{tabular}
\end{table*}

\begin{table*}[]
\caption{Performance comparison between the main baseline SGMSE+ and the proposed FLOWER models at 20 sampling steps. Lower scores indicate better performance.}
\centering
\begin{tabular}{c||c|ccc||c|ccc}
\hline \toprule
Method      & Scenarios                & LSD ($\downarrow$)  & LSD-H ($\downarrow$) & LSD-L ($\downarrow$) & Scenarios                   & LSD ($\downarrow$)  & LSD-H ($\downarrow$) & LSD-L ($\downarrow$)  \\ \addlinespace[0.4mm]\hline \addlinespace[0.7mm]
Unprocessed & \multirow{3}{*}{Matched} & 4.87 & 5.51  & 4.23  & \multirow{3}{*}{Mismatched} & 4.78 & 5.55  & 4.00  \\ \addlinespace[0.7mm]
SGMSE+       &                          & 2.59 & 3.00  & 2.18  &                             & 3.11 & 3.83  & 2.38  \\ \addlinespace[0.7mm]
 \textbf{FLOWER}      &                          & \textbf{2.37} & \textbf{2.69}  & \textbf{2.05}  &                             & \textbf{2.80} & \textbf{3.35}  & \textbf{2.24}  \\\addlinespace[0.4mm] \Xhline{1.1pt}
\end{tabular}
\end{table*}
\section{Experiments}
\subsection{Datasets}
The experiments were conducted on the open vocabulary task of the WSJ dataset \cite{wsj}, a corpus of English reading speech.
The dataset comprised 37,416 utterances for training, 503 for validation, and 333 for testing. For noise distortion, we added CHiME-4 noise to the WSJ dataset. The CHiME-4 noise dataset \cite{chime} included recordings from street, café, bus, and pedestrian environments, with the SNR levels for each utterance randomly selected between 0 and 20 dB.
Reverberation effects were simulated by convolving the speech signal with a room impulse response generated by a Pyroomacoustics engine. 
The simulated rooms had dimensions ranging from 5 to 10 m in length and width, and heights from 2 to 6 m. 
The reverberation time (RT60) for these rooms ranged from 0.3 to 0.9 s.
Bandwidth degradation was achieved using various low-pass filters, including Butterworth, Bessel, Chebyshev, and elliptic filters. The cut-off frequencies for these filters were randomly selected between 2 and 4 kHz, creating a variety of bandwidth degradation effects.

We constructed two test sets for evaluation: ``matched" and ``mismatched." The matched test set was generated from the WSJ+CHiME4 dataset, following the same distortion selection methodology used during training. 
In contrast, the mismatched test set was created using the VCTK+DEMAND dataset \cite{vctk}, which was not used during training, by applying the same reverberation and bandwidth degradation distortions.

\subsection{Experimental settings}
\subsubsection{Evaluation metrics.}
We evaluated the speech restoration performance using various assessment metrics. To assess speech quality, we employed the wide-band perceptual evaluation of speech quality (PESQ) score \cite{pesq}. 
Furthermore, we examined speech signal distortion, background noise, and overall quality using three mean opinion score predictors: CSIG, CBAK, and COVL. CSIG measures signal distortion, CBAK assesses background noise intrusiveness, and COVL evaluates overall signal quality \cite{composite}.
For waveform reconstruction, we utilized a scale-invariant signal-to-distortion ratio (SI-SDR). The speech-to-reverberation modulation energy ratio (SRMR) metric \cite{srmr} was used to evaluate the effectiveness of speech dereverberation. For BWE, we measured the performance using the log-spectral distance (LSD), which was evaluated separately for the high band (4-8 kHz) and low band ( 0-4 kHz ), denoted as LSD-H  and LSD-L, respectively. Higher values indicate better performance for all metrics except for LSD.

\subsubsection{Implementation details.} 
As a main baseline, we employed SGMSE+ \cite{sgmse++}. Originally designed for single distortion tasks, it can handle denoising, dereverberation, and bandwidth extension using the same network architecture, making it suitable for addressing multiple distortions simultaneously.
The FLOWER framework was trained under the same settings as the baseline without task-specific hyperparameter tuning. To ensure a fair comparison, we trained all models for 300 epochs using an NVIDIA A100 GPU with a batch size of 16. For optimization, we employed an Adam optimizer \cite{adam} with a learning rate of $1 \times 10^{-4}$.
We utilized 16 kHz audio data by converting it into a complex-valued short-time Fourier transform representation, with a window size of 510 samples and a hop size of 128 samples, employing a Hann window.
For the NF network, we used an open-source code and modified it to fit our approach. The detailed parameters of the additional modules for the FLOWER approach are described in the caption of Figure 2. Further implementation details and experimental results for other comparative models in GSR tasks are provided in the Appendix for reference.
\begin{figure}
  \centering
  \includegraphics[ width=\linewidth]{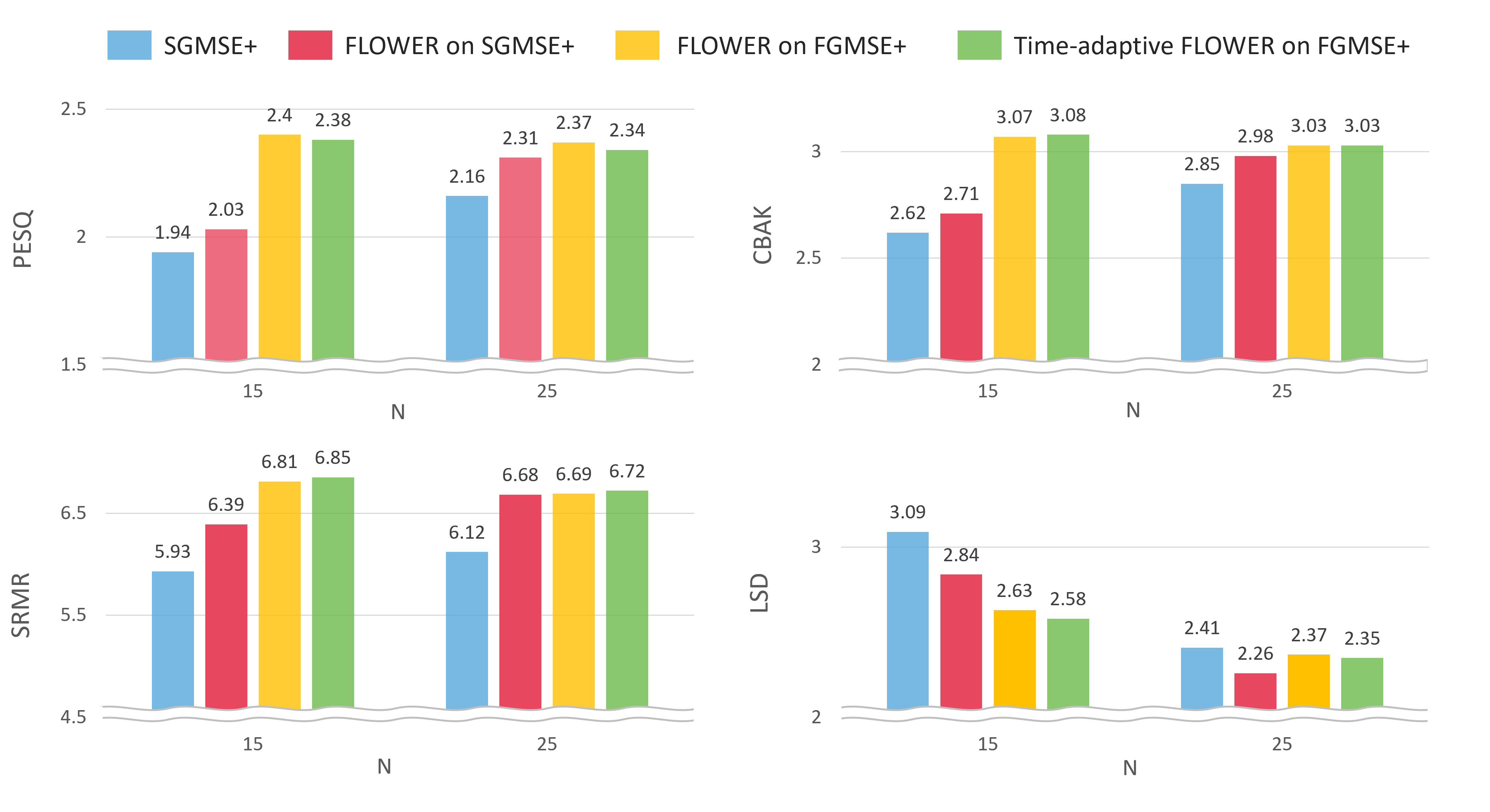}
  \caption{Performance comparison between the SGMSE+ model and the FLOWER method on SGMSE+ and FGMSE+ according to the number of sampling steps ($N$=15, 25) under the ``matched" scenario, respectively. The $x$-axis represents the number of sampling steps ($N$), while the $y$-axis indicates the metric scores. Higher scores are better for PESQ, CBAK, and SRMR, while lower scores are better for LSD. Colors: \textcolor{Gblue}{SGMSE+ (blue)}, \textcolor{Gred}{FLOWER on SGMSE+ (red)}, \textcolor{Gyellow}{FLOWER on FGMSE+ (yellow)}, \textcolor{Ggreen}{Time-adaptive FLOWER on FGMSE+ (green)}.}
\end{figure}
\subsection{Performance Analysis of the FLOWER Approach}
We compared the proposed FLOWER approach with the baseline SGMSE+ model on the WSJ+CHiME4 dataset, which contains multiple distortions. The results are summarized in Tables 1 and 2, and illustrated in Figure 3.
Table 1 evaluates performance in terms of speech quality, noise removal, and dereverberation using the PESQ, CSIG, CBAK, COVL, SRMR, and SISDR metrics.
Table 2 focuses on bandwidth generation performance, using the LSD score, split into LSD-H and LSD-L, to assess high-band reconstruction quality and low-band distortion.
The task involves restoring bandwidths randomly degraded to 2-4 $kHz$ back to the original 8 $kHz$, necessitating evaluation in both bands.
Figure 3 provides a visual comparison of restoration results.
Based on the evaluation results, several conclusions can be drawn:

\noindent i) From Table 1, FLOWER consistently outperformed the main baseline across all six metrics, particularly in noise and reverberation removal. The superior performance in mismatched scenarios indicated FLOWER's robust generalization capabilities.

\noindent ii) As presented in Table 2, FLOWER achieved better results in bandwidth reconstruction for both LSD-H and LSD-L, particularly in mismatched scenarios. The lower LSD scores highlighted its enhanced ability to restore high-band frequencies while minimizing low-band distortion.

\noindent iii) In Figure 3, FLOWER shows clear advantages in noise and reverberation removal compared to the baseline. Notably, in high-band frequencies (highlighted in red boxes), the baseline model introduced more distortions relative to the original spectrum, whereas FLOWER effectively mitigated these issues.

These results demonstrated the efficacy of the FLOWER approach in producing high-quality speech samples.

\subsection{Performance and Efficiency Comparison}
We extended our approach to the FGMSE+ model to improve inference efficiency, focusing on reducing the number of sampling steps $N$. 
Typically, reducing $N$ degrades speech quality, but our FLOWER approach on the FGMSE+ seeks to maintain high performance with fewer steps.
Figure 4 presents the efficiency and performance using four metrics: PESQ for speech quality, CBAK for noise-removal efficiency, SRMR for dereverberation performance, and LSD to evaluate the extent of band restoration.

``FLOWER on SGMSE+” significantly outperformed the baseline ``SGMSE+", indicating better restoration ability of distortions. ``FLOWER on FGMSE+" further improved PESQ, CBAK, and SRMR scores, maintaining higher speech quality even with fewer sampling steps. Specifically, ``FLOWER on FGMSE+" at 15 steps surpassed ``SGMSE+" at 25 steps, demonstrating the efficiency of the FM model.
For example, in the PESQ score, ``SGMSE+" reached 2.16 at 25 steps, while ``FLOWER on FGMSE+" achieved a higher score of 2.4 with only 15 steps. Similar trends were observed across other metrics, such as CBAK and SRMR, where ``FLOWER on FGMSE+" consistently outperformed ``SGMSE+" with fewer sampling steps.
These results highlighted the notable efficiency and effectiveness of the FLOWER approach, which not only reduced the required number of sampling steps but also consistently enhanced performance across key metrics. 
Although ``FLOWER on FGMSE+" shows improvement over ``SGMSE+" in LSD scores, it did not outperform ``FLOWER on SGMSE+". This is due to the FM model's stronger ability to remove residual distortions, including those in high-band frequencies, as observed in Figure 3.

To enhance band reconstruction, we applied time-adaptive Gaussian guidance in ``Time-adaptive FLOWER on FGMSE+". This method, which adjusts conditioning information inversely proportional to the time step, improved LSD scores without excessive high-band removal. Figure 3 red boxes confirmed this, showing that the time-adaptive model produced spectrograms more closely resembling the original.

\section{Conclusion and Discussion}
In this work, we propose a novel conditioning approach, FLOWER, which integrates Gaussian guidance into generative frameworks, significantly improving speech restoration performance by effectively handling multiple distortions. Through extensive evaluation, we demonstrated its superiority over baseline methods, showcasing enhanced noise removal, dereverberation, and band reconstruction capabilities. The adaptability of our approach to various scenarios, including matched and mismatched datasets, underscored its robustness and generalization ability. By reducing the number of sampling steps and incorporating time information, our method achieved efficient and comprehensive speech restoration. Overall, FLOWER presented a versatile and effective solution for real-world speech restoration challenges, offering high-quality speech outputs.
However, we acknowledge that our task was limited to three types of distortions.  In future work, we plan to address a broader range of distortions and explore more effective restoration methods.


\bibliography{example_paper}
\bibliographystyle{icml2025}

\newpage
\appendix
\onecolumn
\section{Comparative Models for General Speech Restoration}
We evaluated the proposed FLOWER approach against several comparative models designed for speech enhancement and general speech restoration tasks, focusing on generative models capable of addressing noise removal or managing multiple distortions simultaneously.
The application of diffusion models in speech enhancement began with DiffuSE \cite{diffuse} and SGMSE \cite{sgmse}, which introduced foundational generative approaches for this domain.
Building on these, CDiffuSE extended DiffuSE by incorporating a conditional diffusion framework, enhancing its ability to handle more complex conditions.
SGMSE+ improved upon SGMSE by adopting the NCSN++ architecture as its backbone network, resulting in significant performance gains.
While originally proposed for single-distortion tasks such as denoising, dereverberation, and BWE, SGMSE+ has demonstrated effectiveness in addressing multiple distortions concurrently, making it a strong main baseline for general speech restoration. 
UniverSE introduced a discriminative conditioning network that works alongside diffusion networks. By extracting enhanced features through a discriminative network, it provides guidance to score-based diffusion models via joint training.
Building on this foundation, UniverSE++ further refined the framework by applying adversarial loss and structural modifications, achieving notable improvements in restoration quality.

\section{Experimental Results}
\subsection{Quantitative Comparison}
\begin{table*}[h]
\centering
\caption{Performance comparison with several comparative models under ``matched” scenarios. Higher scores indicate better performance for PESQ, CSIG, CBAK, COVL, and SRMR, while lower scores indicate better performance for LSD.}
\label{tab:performance_comparison}
\resizebox{\linewidth}{!}{%
\begin{tabular}{cccccccc}
\hline \toprule
\textbf{Model} & \textbf{PESQ} ($\uparrow$) & \textbf{CSIG} ($\uparrow$) & \textbf{CBAK} ($\uparrow$) & \textbf{COVL} ($\uparrow$) & \textbf{SRMR} ($\uparrow$) & \textbf{SI-SDR} ($\uparrow$) & \textbf{LSD} ($\downarrow$) \\ \addlinespace[0.4mm] \toprule \addlinespace[0.7mm]
Unprocessed         & 1.61 & 1.80 & 2.21 & 1.65 & 3.96 & 2.54 & 4.87  \\ \addlinespace[0.7mm]
CDiffuSE            & 1.52 & 2.83 & 2.20 & 2.22 & 7.20 & -1.31 & 3.47 \\
\addlinespace[0.7mm]
SGMSE+              & 2.10 & 3.52 & 2.78 & 2.87 & 6.08 & 7.14 & 2.59 \\
\addlinespace[0.7mm]
UniverSE            & 1.97 & 3.37 & 2.71 & 2.72 & 6.79 & 3.17 & 2.68 \\
\addlinespace[0.7mm]
UniverSE++          & 2.12 & 3.27 & 2.85 & 2.75 & 6.13 & 5.09 & 2.67 \\
\addlinespace[0.7mm]
 \textbf{FLOWER}   & 2.23 & 3.62 & 2.90 & 3.00 & 6.61 & 8.05 & 2.37 \\ \toprule
\end{tabular}%
}
\end{table*}

\begin{table*}[h]
\centering
\caption{Performance comparison with several comparative models under ``mismatched” scenarios.}
\label{tab:performance_comparison2}
\resizebox{\linewidth}{!}{%
\begin{tabular}{cccccccc}
\hline \toprule
\textbf{Model} & \textbf{PESQ} ($\uparrow$) & \textbf{CSIG} ($\uparrow$) & \textbf{CBAK} ($\uparrow$) & \textbf{COVL}  ($\uparrow$) & \textbf{SRMR} ($\uparrow$) & \textbf{SI-SDR} ($\uparrow$) & \textbf{LSD} ($\downarrow$) \\ \addlinespace[0.4mm] \toprule
\addlinespace[0.7mm]
Unprocessed         & 1.78 & 1.80 & 2.25 & 1.73 & 5.91 & 2.58 & 4.78 \\
\addlinespace[0.7mm]
CDiffuSE            & 1.56 & 2.64 & 2.25 & 2.12 & 9.54 & -0.63 & 3.29 \\
\addlinespace[0.7mm]
SGMSE+              & 1.82 & 2.83 & 2.70 & 2.37 & 8.57 & 6.63 & 3.11 \\
\addlinespace[0.7mm]
UniverSE            & 1.65 & 2.64 & 2.47 & 2.18 & 9.08 & 1.71 & 2.93 \\
\addlinespace[0.7mm]
UniverSE++          & 1.81 & 2.97 & 2.67 & 2.43 & 8.17 & 3.62 & 2.73 \\
\addlinespace[0.7mm]
 \textbf{FLOWER}   & 1.95 & 2.99 & 2.81 & 2.51 & 9.34 & 7.55 & 2.80 \\ \hline
\toprule
\end{tabular}%
}
\end{table*}

\subsection{Implementation Details}
\begin{itemize}
    \item CDiffuSE \cite{cdiffu} \\
    We trained the CDiffuSE model using the large configuration provided at \url{ https://github.com/neillu23/CDiffuSE }, adhering to most settings specified in the original paper for the large model. The model was trained for 300,000 iterations with a batch size of 15, employing an early stopping scheme. The training utilized 200 diffusion steps to ensure performance aligned with the intended configuration.
    \item SGMSE+ \cite{sgmse+} \\
    The implementation followed the details outlined in Section 5.2.2 of this paper. The model comprises 65 M trainable parameters and was trained without task-specific hyperparameter tuning. The official implementation of SGMSE+ is available at \url{ https://github.com/sp-uhh/sgmse.git}, ensuring reproducibility and alignment with prior research.
    \item UniverSE \cite{universal}\\
    UniverSE was trained using the code in \url{ https://github.com/line/open-universe}. The training process involved 1,500,000 steps with a batch size of 20, utilizing two GPUs. The model included 46.4 M trainable parameters.
    \item UniverSE++ \cite{universepp}\\
    UniverSE++ was evaluated using the same codebase as UniverSE, available at \url{ https://github.com/line/open-universe }. The model is trained for 1,500,000 training steps with a batch size of 20 on two GPUs. The model included 84.2 M trainable parameters.
    
    \item FLOWER (Ours)\\
    The model was trained using the generative network and the NF network available at \url{https://github.com/jaywalnut310/vits}. The model comprises 65.5 M trainable parameters, with detailed parameter settings described in Section 5.2.2. 
\end{itemize}

\subsection{Qualitative Analysis}

\begin{figure*}[t]
\centering
\includegraphics[width=\linewidth]{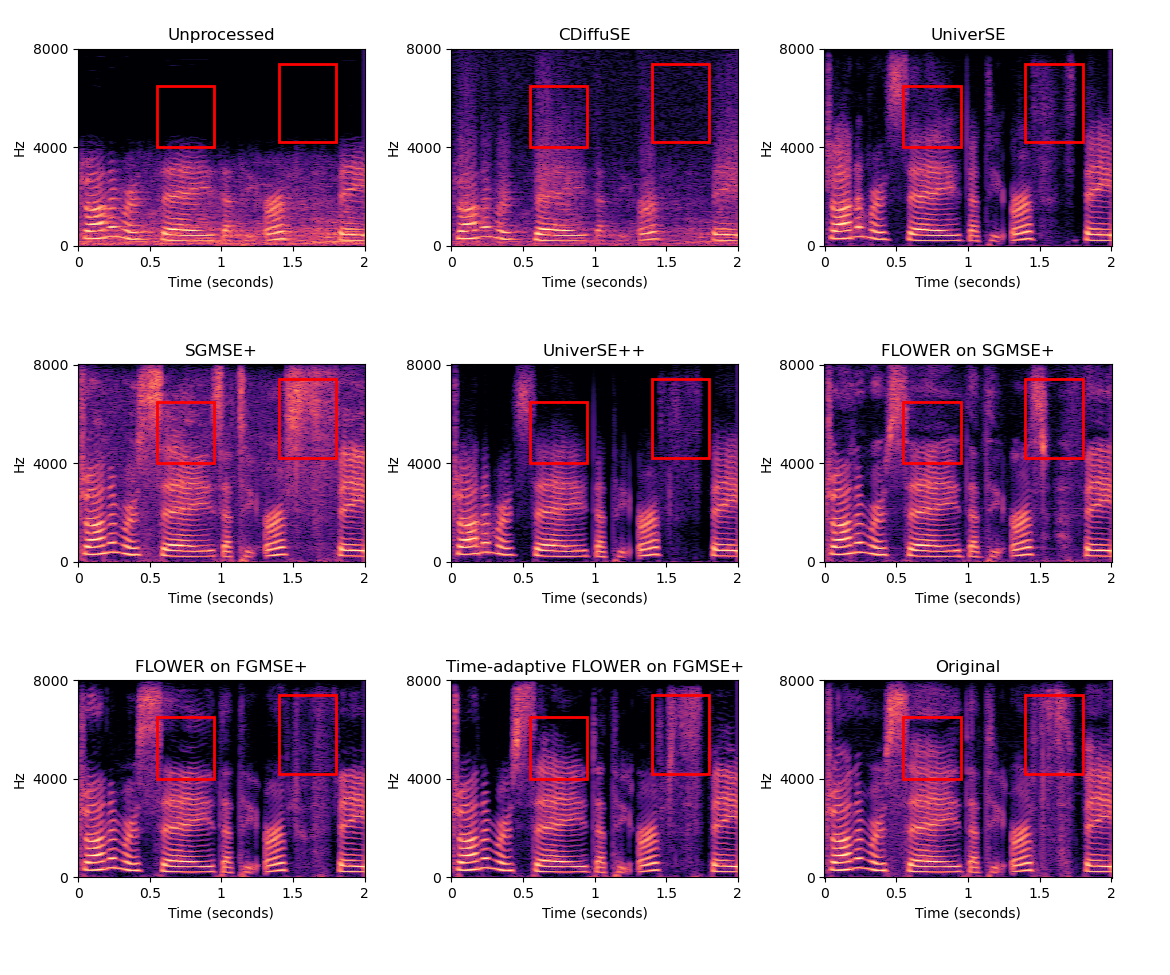}
\caption{Spectrogram analysis comparing the proposed method to baseline models. The results demonstrate the efficacy of the proposed approach in preserving spectral details and mitigating distortions. Notably, the ``Time-adaptive FLOWER on FGMSE+" model further alleviated residual reverberations present in the original signals.}
\end{figure*}

\begin{figure*}[t]
\centering
\includegraphics[width=\linewidth]{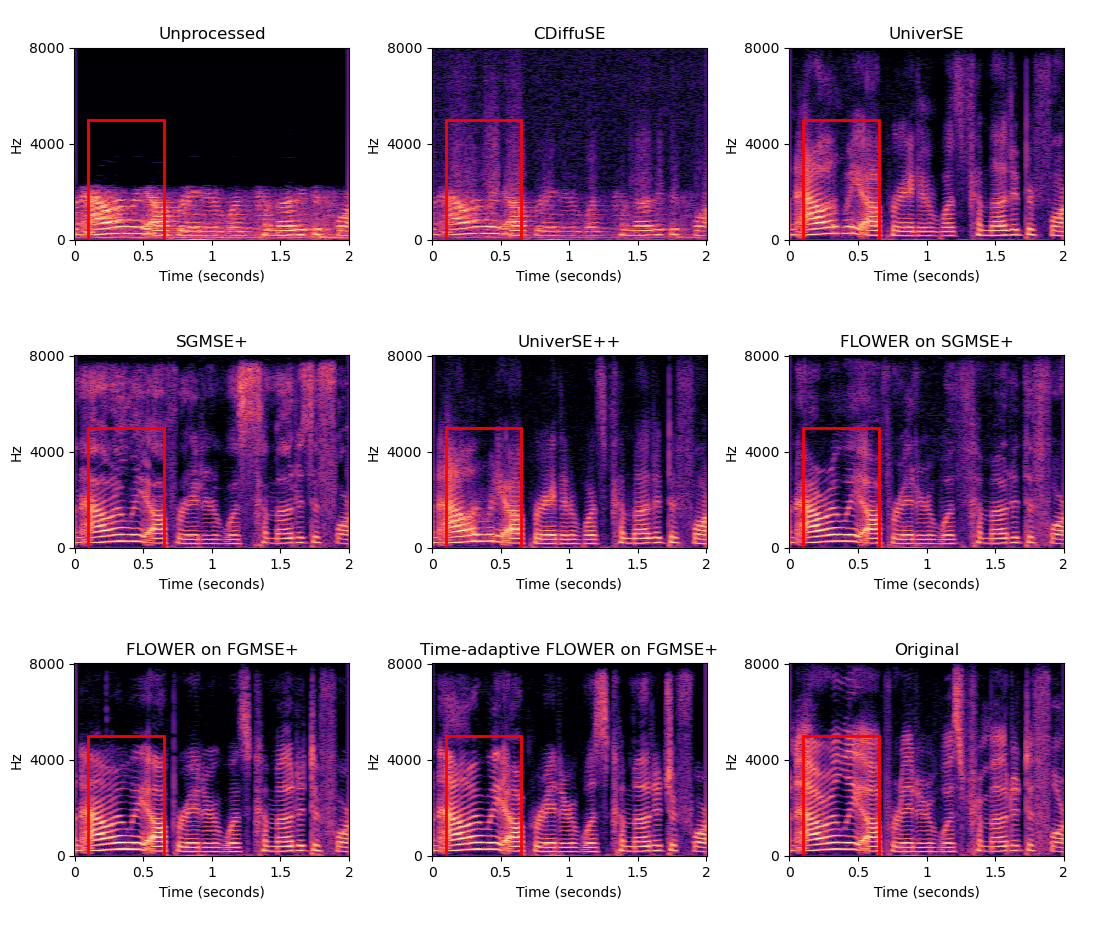}
\caption{Spectrogram analysis comparing the proposed method to baseline models.}
\end{figure*}


\end{document}